\documentclass[%
 reprint,
superscriptaddress,
amsmath,amssymb,
aps,
]{revtex4-2}

\newcommand{\apjl}{Astrophys. J. Lett.}   
\newcommand{\apjs}{Astrophys. J. Suppl. Ser.}   
\newcommand{\mnras}{Mon. Not. R. Astron. Soc.}   
\newcommand{\nastro}{Nat. Astron.} 
\newcommand{\physrep}{Phys. Rep.}   
\newcommand{\sci}{Science} 

\newcommand{\npa}{Nucl. Phys. A}
\newcommand{\plb}{Phys. Lett. B}
\newcommand{\epja}{Eur. Phys. J. A}

\newcommand{\pr}{Phys. Rev.}

\usepackage[normalem]{ulem}
\usepackage{xcolor}
\usepackage[symbol*]{footmisc}
\setfnsymbol{wiley}
\usepackage[utf8]{inputenc}
\usepackage{graphicx}
\usepackage{dcolumn}
\usepackage{bm}
\usepackage{tabularx}
\usepackage{amsmath}
\usepackage{array}
\usepackage{comment}
\usepackage{verbatim}
\usepackage{siunitx}
\usepackage[colorlinks=true,
linkcolor=blue,anchorcolor=blue,citecolor=blue,
hypertexnames=false
]{hyperref}
\usepackage{wrapfig}

\DeclareRobustCommand{\sout}{\bgroup\markoverwith{\textcolor{red}{\rule[.5ex]{2pt}{1pt}}}\ULon}

\makeatletter
\newcommand\footnoteref[1]{\protected@xdef\@thefnmark{\ref{#1}}\@footnotemark}
\makeatother

\begin{document}

\title{Determination of the Equation of State from Nuclear Experiments and Neutron Star Observations}


\author{Chun~Yuen~Tsang}
\affiliation{Facility for Rare Isotope Beams, Michigan State University, East Lansing, Michigan 48824, USA}
\affiliation{Department of Physics and Astronomy, Michigan State University, East Lansing, Michigan 48824, USA}

\author{ManYee~Betty~Tsang\footnote{Corresponding author. Email: tsang@frib.msu.edu}}
\affiliation{Facility for Rare Isotope Beams, Michigan State University, East Lansing, Michigan 48824, USA}
\affiliation{Department of Physics and Astronomy, Michigan State University, East Lansing, Michigan 48824, USA}

\author{William~G.~Lynch}
\affiliation{Facility for Rare Isotope Beams, Michigan State University, East Lansing, Michigan 48824, USA}
\affiliation{Department of Physics and Astronomy, Michigan State University, East Lansing, Michigan 48824, USA}

\author{Rohit~Kumar}
\affiliation{Facility for Rare Isotope Beams, Michigan State University, East Lansing, Michigan 48824, USA}

\author{Charles J. Horowitz}
\affiliation{Facility for Rare Isotope Beams, Michigan State University, East Lansing, Michigan 48824, USA}
\affiliation{Center for the Exploration of Energy and Matter and Department of Physics, Indiana University, Bloomington, IN 47405, USA}

\date{\today}





\begin{abstract}
With recent advances in neutron star observations, major progress has been  made in determining the pressure of neutron star matter at high density. This pressure is constrained by the neutron star deformability, determined from gravitational waves emitted in a neutron-star merger, and measurements of radii of two neutron stars, using a new X-ray observatory on the International Space Station.
Previous studies have relied on nuclear theory calculations to provide the equation of state at low density. Here we use a combination of 15 constraints composed of three astronomical observations and twelve nuclear experimental constraints that extend over a wide range of densities. Bayesian Inference is then used to obtain a comprehensive nuclear equation of state. This data-centric result provides benchmarks for theoretical calculations and modeling of nuclear matter and neutron stars. Furthermore, it provides insights
on the composition of neutron stars and their cooling via neutrino radiation. 
\end{abstract}

\maketitle

\section{Introduction}
With masses that can be larger than twice the mass of the sun and with radii of approximately 13 km, neutron stars (NS) and their observed properties raise some compelling questions \cite{lattimer2004physics}. Are pions, strange particles or quark matter important for understanding NS matter and its internal pressure, which supports a NS and prevents it from collapsing into a black hole? If that matter is nucleonic, what roles do repulsive short-range three-neutron or four-neutron forces play in supporting the star? 
Understanding the connections between the pressure and the composition and the structure of the stellar matter clearly constitutes a key objective for both astrophysics and nuclear physics \cite{dietrich2020}. To better understand the connections between nuclei and neutron stars \cite{hen2021}, we combine nuclear measurements and astrophysical observations to obtain an equation of state (EOS) of nuclear matter. 

The breakthrough in equation of state research in the past 5 years has been in Astronomy. First the era of Multi-Messenger astronomy, involving simultaneous observations of a merging binary NS system with a wide array of astronomical instruments \cite{abbott2017}, has enabled more detailed studies of merging neutron stars. Then, these two merging neutron stars (GW170817) observed by the Gravitational-Wave Observatories LIGO/Virgo Collaboration provided information on the deformability of the neutron stars \cite{abbott2017gw170817}. Next, the measurements of the radii of pulsars  by the Neutron Star Interior Composition Explorer (NICER) \cite{riley2019,miller2019,riley2021,miller2021} provided constraints on the NS mass-radius correlation. Both the deformability and mass-radius relationship constrain the pressure-density relationships or EOS of NS matter above twice ``saturation" density, $n > 2n_0$ \cite{legred2021,abbott2018gw170817}, where $n_0 \approx \SI{0.16}{nucleons/fm^3}$ ($\approx\SI{2.6e14}{g/cm^3}$) is the density at the core of any heavy nucleus. To augment these astrophysical constraints, we add constraints from nuclear physics experiments  \cite{danielewicz2002determination,le2016constraining,dutra2012skyrme,zhang2015electric,adhikari2021accurate,reed2021implications,brown2013constraints,kortelainen2012nuclear,danielewicz2017symmetry,tsang2009constraints,morfouace2019constraining,estee2021probing,cozma2018feasibility,RUSSOTTO2011471,russotto2016results} obtained in the past two decades. When combined, these constraints provide a consistent description of the EOS that describes nuclear matter and neutron stars at densities of $0.5n_0 < n < 3n_0$.  

 A Bayesian analysis framework is used to infer the EOS parameters and quantify the uncertainties in the results. 
In addition to describing the data from nuclear physics experiments, we also use a neutron star model that employs the same EOS function to calculate neutron star properties so that these predictions can be compared to astronomical observations. The resulting EOS serves as a benchmark for the microscopic nuclear theory and provides insights into the nature of strongly interacting matter in the outer core of the NS, its composition, and the onset of direct Urca cooling processes. In a different context, a comprehensive EOS would also yield insights into the collapse of supernovae and emission of neutrinos.

\section{Equation of State Function} 
We assume nucleonic matter in nuclei and in neutron stars at zero temperature share a common nuclear EOS that can be chosen to be the energy per nucleon as a function of density, $\epsilon(n)$ and obtain the pressure by differentiating $\epsilon(n)$ with respect to density. We parameterize $\epsilon(n_n,n_p)$ as a function of the neutron and proton number densities $n_{n}$ and $n_{p}$, where $n$=$n_{n}$+$n_{p}$ is the total nucleonic density and $\delta=\frac{n_{n}-n_{p}}{n}$ is the asymmetry parameter.   
To describe the evolution of the EOS with density and asymmetry, $\delta$, $\epsilon(n_n,n_p)$ =$\epsilon_{SNM}(n)$+$S(n)\delta^2$ consists of two terms where $\epsilon_{SNM}(n)$ is the energy per nucleon for symmetric nuclear matter with $\delta$=0 and $S(n)\delta^2$ is the symmetry energy term needed when $\delta\neq 0$.  This allows the EOS to describe nuclear matter with any neutron-proton composition. Such a division is useful because the asymmetry $\delta$ can be experimentally controlled to enhance  the sensitivity to either the symmetric matter EOS or to the symmetry energy term. Detailed description of the equations of state used in the Bayesian analysis can be found under Methods.

\section{Constraints}

\begin{figure}
\resizebox{0.5\textwidth}{!}
{\includegraphics[trim={4cm 0 0 0},clip]{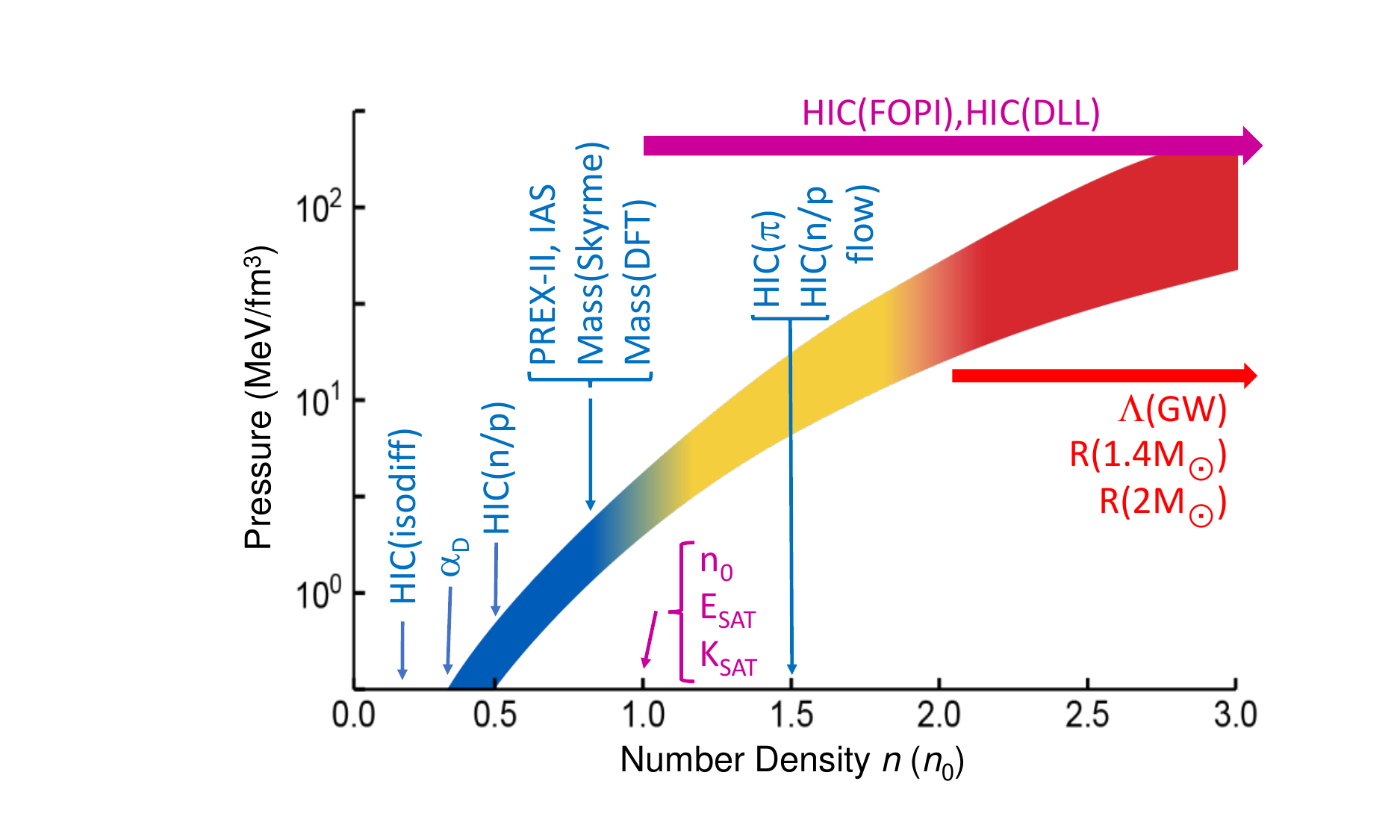}}
\vspace{-0.3in}
\caption{\textbf{Constraints from nuclear experiments and astronomy observations with their corresponding sensitive densities.} The red horizontal arrow indicates the high density region ($>2 n_0$) probed by neutron star observations. The magenta text indicates chosen values  of $E_{SAT}\approx\SI{-16}{MeV}$ for the symmetric matter EOS located at the saturation density  $n_0\approx \SI{0.16}{fm^{-3}}$ following \cite{drischler2019, drischler2016} and a symmetric matter constraint on $K_{SAT}$  obtained from Giant Monopole Resonance (GMR) studies \cite{dutra2012skyrme}. The purple horizontal arrow at the top of the figure indicates the range of density from symmetric matter constraints coming from heavy-ion collision studies \cite{danielewicz2002determination,le2016constraining}. The remaining terrestrial constraints for nuclear symmetry energy or symmetry pressure are labelled in blue and positioned at their sensitive  densities \cite{decoding2022}. The colored band indicates the approximate dependence of pressure on density for pure neutron matter from the findings of current work.}
    \label{fig:all_constraints}
\end{figure}

Most experimental and astronomical observables are sensitive to the EOS over a limited range of density.  For example, radii, masses and deformabilities of neutron stars largely reflect the pressure inside the neutron stars 
at densities of $n>2n_0$ \cite{legred2021,Miller_2020}. On the other hand, nuclear masses and neutron skins of neutron-rich nuclei are mostly sensitive to the nuclear matter EOS at (2/3)$n_{0}$ \cite{decoding2022}.
In addition to these two density regions, observables constructed from the distribution patterns of particles 
emitted in nucleus-nucleus or heavy ion collisions (HIC) have provided constraints on the EOS at densities range $0.22 < n/n_0< 4.6$ 
\cite{danielewicz2002determination,le2016constraining,tsang2004isospin,tsang2009constraints,tsang2012constraints,morfouace2019constraining}.
The diversity of the  available data is given in Table~\ref{tab:constraints} and also illustrated in Fig.~\ref{fig:all_constraints} which shows the effective density region of mostly nuclear physics constraints up to about 3$n_0$.  Constraints prefaced with ``HIC" are obtained from heavy ion collision experiments. Brief description of each of the constraints will be discussed below.  For astrophysics, the relevant density range is above 2$n_0$.    
Recall that the nuclear equation of state is parameterized into two terms, the symmetric matter term and the symmetry energy term that accounts for the imbalance of neutrons and protons. We separate the nuclear physics constraints into ones that are sensitive to the symmetry energy and others that probe  the symmetric matter EOS.

\begin{table*}

\caption{\textbf{List of Constraints used in Bayesian Analysis}. The NICER constraints marked with * are given half weight in the Bayesian analysis as the 4 constraints listed here come from two measurements analyzed by two different groups.}
\label{tab:constraints}
\begin{tabular*}{\textwidth}{@{\extracolsep{\fill}}lcccc}
\hline
Symmetric matter  &&&& \\
Constraints & $n$ (fm$^{-3}$) & P$_\text{SNM}$ (MeV/fm$^3$) & $K_\text{SAT}$ (MeV) & Ref. \\
\hline
HIC(DLL)  & 0.32             & $10.1\pm3.0$                 &                      &\cite{danielewicz2002determination} \\
HIC(FOPI)  & 0.32                & $10.3\pm2.8$                 &                      &\cite{le2016constraining} \\
GMR          & 0.16              &                              & $230\pm30$           & \cite{dutra2012skyrme} \\ \\
\hline
\hline
Asymmetric matter  &&&& \\
Constraints & $n$ (fm$^{-3}$) & S($n$) (MeV)      & P$_\text{sym}$ (MeV/fm$^3$) & Ref. \\ 
\hline
Nuclear structure &&&&\\
$\alpha_D$    & 0.05           & $15.9\pm1.0$ &                                        &~\cite{zhang2015electric}        \\
PREX-II       & 0.11             &              & $2.38\pm0.75$           &~\cite{adhikari2021accurate,reed2021implications,decoding2022}\\
\\
Nuclear masses &&&&\\ 
Mass(Skyrme)  & 0.101$\pm$0.005           & $24.7\pm0.8$ &                                          &~\cite{brown2013constraints,decoding2022}      \\
Mass(DFT) &  0.115$\pm$0.002               & $25.4\pm1.1$ &                                             &~\cite{kortelainen2012nuclear,decoding2022}  \\
IAS           & 0.106$\pm$0.006           & $25.5\pm1.1$ &                                             &~\cite{danielewicz2017symmetry,decoding2022}     \\
\\
Heavy-ion collisions \\ 
HIC(Isodiff)  & 0.035$\pm$0.011          & $10.3\pm1.0$ &                                             &~\cite{tsang2009constraints,decoding2022}    \\
HIC(n/p ratio) & $0.069\pm0.008$\          & $16.8\pm1.2$ &                                         &~\cite{morfouace2019constraining,decoding2022}\\

HIC($\pi$)    & 0.232$\pm$0.032           & $52\pm13$    & $10.9\pm8.7$                         &~\cite{estee2021probing,decoding2022}    \\

HIC(n/p flow) & 0.240           &              & $12.1\pm8.4$          &~\cite{cozma2018feasibility,RUSSOTTO2011471,russotto2016results,decoding2022} \\
\\
\hline
\hline
\end{tabular*}

\begin{tabular*}{\textwidth}{@{\extracolsep{\fill}}lcccc}
Astronomical    &&&& \\
Constraints & $M(\odot)$ & R (km) & $\Lambda$ & Ref.\\
\hline 
LIGO & 1.4 & & $190_{-120}^{+390}$ &~\cite{abbott2018gw170817} \\ \vspace{0.2mm}
*Riley ~PSR J0030+0451& $1.34^{+0.15}_{-0.16}$ & $12.71^{+1.14}_{-1.19}$ & &~\cite{riley2019} \\\vspace{0.2mm}
*Miller PSR J0030+0451& $1.44^{+0.15}_{-0.14}$ & $13.02^{+1.24}_{-1.06}$ & &~\cite{miller2019} \\\vspace{0.2mm}
*Riley ~PSR J0740+6620& $2.07^{+0.07}_{-0.07}$ & $12.39^{+1.30}_{-0.98}$ & &~\cite{riley2021}\\\vspace{0.2mm}
*Miller PSR J0740+6620 & $2.08^{+0.07}_{-0.07}$ & $13.7~^{+2.6}_{-1.5}$ & &~\cite{miller2021}
 \\
\hline

\end{tabular*}
\end{table*}
\subsection{Symmetric Matter Constraints}
 Certain properties of the symmetric nuclear matter such as the saturation density, $n_0$, and saturation energy, $E_{SAT}$ are reasonably well determined. We adopt the values of $n_0 \approx$ \SI{0.16}{fm^{-3}} and $E_{SAT}\approx$ \SI{-16}{MeV} from \cite{drischler2019, drischler2016}.  
 For incompressibility
parameter, $K_{SAT}$, we use the values of \SI[separate-uncertainty=true, multi-part-units=single]{230 \pm 30}{MeV} following Ref.~\cite{dutra2012skyrme}. Even though these three parameters are labelled in Fig.~\ref{fig:all_constraints} at $n_0$, only the Taylor expansion coefficient, $K_{SAT}$, is used as a constraint in the Bayesian analysis. 

\begin{figure*}
\vspace{-0.3in}
\includegraphics{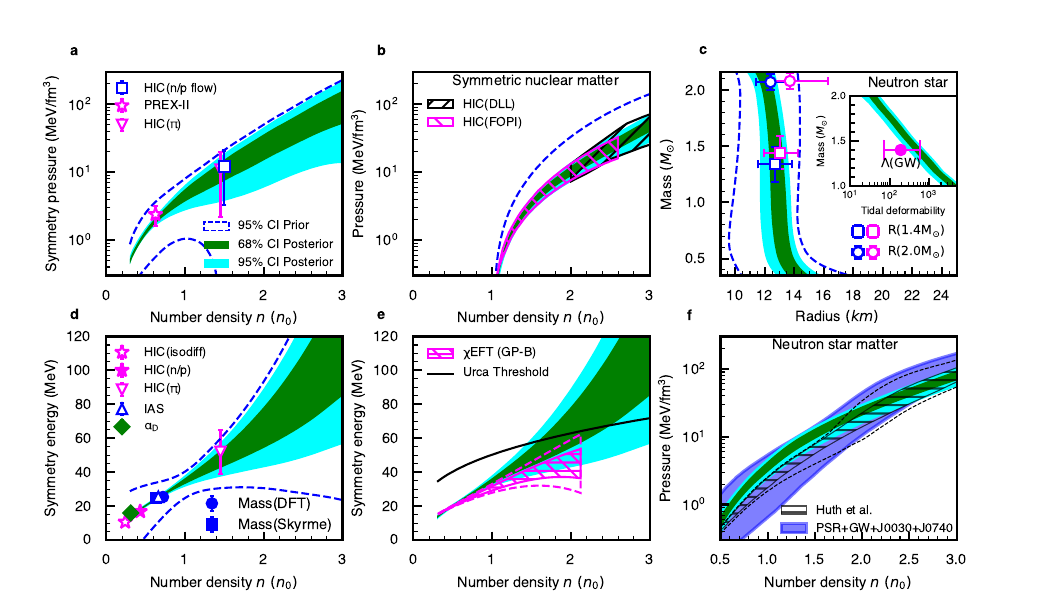}
\caption{\textbf{Constraints on and posteriors of the nuclear equation of state and neutron star.} All symbols in these panels are data discussed in the text and listed in Table~\ref{tab:constraints}. The error bars reflect 1-$\sigma$ standard deviation unless specified. The blue dashed lines are 95\% CI boundaries of the prior distributions with an initial sample size of 1.5M. The boundaries of the dark green and light blue shaded regions represent 68\% and 95\% CI boundaries of the posterior distributions respectively using an initial size of 3 M. Panels: (a) Density dependence of the symmetry pressure. (b) Density dependence of symmetric nuclear matter pressure. (c) Mass-Radius correlation plot. Symbols are data from two independent analysis of two pulsers measured by NICER. The insert shows the mass-deformability correlation plot.  The symbol represents the experimental deformability value for $1.4M_\odot$ with error bars reflecting 90\%CI. (d) Density dependence of the symmetry energy. (e) Density dependence of the symmetry energy compared to $\chi$EFT calculations (magenta). The onset of Urca cooling occurs above the solid black line.  (f) Density dependence of the pressure for neutron star matter. The purplish-blue area is the EOS obtained in Ref. \cite{legred2021} based on astronomical constraints using only non-parametric EOS. The hatched area is from Ref. \cite{nature2022} with astronomical and high density heavy ion collision constraints at high energy and  $\chi$EFT as priors at low density.}
    \label{fig:nuclear}
\end{figure*}

Above saturation density, measurements of 
collective flow from energetic Au+Au
collisions have constrained the EOS for symmetric matter at densities ranging from $n_0$ to 4.6$n_0$ \cite{le2016constraining,danielewicz2002determination}. The density range is indicated by a horizontal magenta arrow on the upper right corner of Fig.~\ref{fig:all_constraints}.  
The pressure-density plot for symmetric matter in Fig.~\ref{fig:nuclear}b shows these two flow contours  
\cite{le2016constraining,danielewicz2002determination}. The black slanted-hatched region (HIC(DLL)) corresponds to the symmetric matter EOS published in Ref. \cite{danielewicz2002determination} from the analysis of $both$ collective transverse and elliptical flow data measured at incident energies of 0.2-10 GeV/u. The extraction and model dependence of the HIC(DLL) contours are discussed in \cite{danielewicz2002determination} and its online supplemental material. The magenta slanted-hatched region (HIC(FOPI)) located at lower densities results from an independent analysis of a different set of elliptical flow measurements at 0.4 to 1.5 GeV/u \cite{le2016constraining}. Known sources of theoretical uncertainty are modeled and the symmetric nuclear matter pressure plotted in panel  b reflects this uncertainty estimation. Both contours agree very well in the region around 2$n_0$ where they overlap.  We take the pressure values (open square symbols) at 2$n_0$ as a constraint in our analyses. 
In view of availability of the new and better measured data \cite{HADES:2020lob,star2021,star2022}, more robust constraints on symmetric matter are expected in the near future. 

\subsection{Symmetry Energy Constraints}
 In the past decade, many studies have been conducted to extract the symmetry energy and its contributions to the pressure (symmetry pressure) \cite{decoding2022} mostly at low densities.  
Precise symmetry energy constraints shown in Fig.~\ref{fig:nuclear}d have been obtained at about (2/3)$n_{0}$  from nuclear masses using density functional theory in two independent analyses involving different nuclei labeled as Mass(Skyrme) \cite{brown2013constraints} and 
Mass(DFT) \cite{kortelainen2012nuclear}. In this density region we also have precise constraints obtained from the energies of isobaric analogue states \cite{danielewicz2017symmetry} indicated by a data point labeled as IAS. Measurements of the dipole polarizability of $^{208}$Pb provide a constraint at a lower density of $n=\SI{0.05}{fm^{-3}}$, which is labeled as $\alpha_{D}$ \cite{zhang2015electric}. Values for the neutron skin thickness and the slope of the symmetry energy, $L(2n_{0}/3)$, at $(2/3)n_{0}$ have been published for the PREX-II experiment from which we obtain the pressure $P(2n_{0}/3)$. This is labeled as PREX-II \cite{adhikari2021accurate,reed2021implications} in Fig.~\ref{fig:nuclear}a.  

Heavy-ion collisions have probed the symmetry energy and pressure at densities far from (2/3)$n_{0}$. At incident energies below 100 MeV per nucleon, lower densities are probed when matter expands after initial impact and compression of the projectile and target. There, experimental observables primarily reflect the symmetry energy at sub-saturation densities with $n <<n_{0}$ \cite{tsang2009constraints,morfouace2019constraining}. Constraints from isospin diffusion in Sn+Sn collisions, labelled as HIC(isodiff) \cite{tsang2009constraints} provide a constraint at $n/n_{0}=0.22\pm0.07$. Ratios of neutron and proton energy spectra in central collisions, labelled as HIC(n/p) provide a constraint on the symmetry energy  at $n/n_{0}=0.43\pm0.05$ \cite{morfouace2019constraining}. 

Higher energy ($>$200 MeV/u) central HIC probe the EOS at $n>n_{0}$. 
The HIC(n/p flow) data point comes from the elliptical collective flows of neutrons and hydrogen nuclei in Au+Au collisions \cite{cozma2018feasibility,RUSSOTTO2011471,russotto2016results}. HIC($\pi$) comes from a recent measurement of the spectral ratios of charged pions in very neutron-rich $^{132}$Sn+$^{124}$Sn collisions and nearly symmetric $^{108}$Sn+$^{112}$Sn collisions \cite{estee2021probing}. The use of ($^{132}$Sn) and ($^{108}$Sn) radioactive beams enhanced the asymmetry variation and the experimental sensitivity to the symmetry energy. The uncertainty of the pion constraint reflects significant uncertainties in the difference between neutron and proton effective masses and in the contributions from non-resonant pion production \cite{estee2021probing}. Extractions and discussions of the model dependent uncertainties in the constraint marked as HIC(FOPI), HIC(n/p) and HIC(n/p flow) can be found in refs. \cite{le2016constraining, estee2021probing, cozma2018feasibility}. Finally, we note that the uncertainties in the sensitive densities listed in Table~\ref{tab:constraints} are too small to significantly influence the final EOS and therefore not used in the Bayesian analysis. 

\subsection{Constraints from astronomical observations}
When two neighboring neutron stars begin to merge, the gravitational field of each neutron star induces a tidal deformation in the other. The influence of the EOS of neutron stars on the gravitational-wave signal during the in-spiral phase is encoded in the dimensionless, neutron-star tidal deformability, $\Lambda$ \cite{abbott2017gw170817}. In the inset of  Fig.~\ref{fig:nuclear}c, the value of $\Lambda$ for $1.4M_\odot$ obtained  from the  binary neutron star merger event of GW170817 \cite{abbott2018gw170817} is plotted. The observation of a later merger event (GW190425) did not improve the accuracy of $\Lambda$ due to poorer observation conditions \cite{Abbott2020}.

Since its operation in 2019, the Neutron Star Interior Composition Explorer (NICER) \cite{riley2019,miller2019,riley2021,miller2021} has measured the radii of two pulsars with very different masses. Figure~\ref{fig:nuclear}c shows values for the radii obtained by Riley et al (Miller et al) for two pulsars, PSR J0030+0451 and PSR J0740+6620 \cite{fonseca2021,cromartie2020} with measured masses of $1.34^{+0.15}_{-0.16}$$M_\odot$ ($1.44^{+0.15}_{-0.14}$$M_\odot$) and $2.07^{+0.07}_{-0.07}$$M_\odot$ ($2.08^{+0.07}_{-0.07}$$M_\odot$), respectively \cite{riley2019,riley2021,miller2019,miller2021}. Since these are independent analysis on the same pulsars, each constraint is given the weight of 0.5 in the analysis.

\section{Description of the neutron star model calculations}  
 
We adopt the NS model described in Ref. \cite{tsang2020impact} to calculate properties of the neutron star, such as the deformability, mass and radius for each EOS in the prior distribution, by solving the Tolman-Oppenheimer-Volkoff (TOV) equation \cite{tolman1939,Oppenheimer1939}. In particular, we focus on the impact of the EOS on the outer core of the neutron star where neutrons comprise the bulk of the matter and inclusion of small admixtures of protons, electrons and muons is required to attain $\beta$-equilibrium. For simplicity, we include no sharp phase transitions to occur within the neutron star core. Both the inner and outer core can be described by the same equation unless the speed of sound in outer core EOS begins to exceed the speed of light. If this occurs, we transition it to the stiffest possible EOS where speed of sound equals to speed of light and the EOS becomes very repulsive. Such transition occurs mostly above $3n_0$. Below the crust-core transition density, we use the crustal EOS of Ref. \cite{crustalEOS} (see \emph{Methods} for details).  Our study provides insights into the EOS at densities of 0.5-3$n_0$ where the current experimental constraints are relevant.

\section{Results}

Bayesian inference is used to simultaneously determine all the parameters used in the EOS constructed from metamodeling (see \emph{Methods} for details). We constrain the Taylor expansion coefficients of the equation of state solely by the 15 discrete constraints obtained from astronomical observations and nuclear physics experiments. These constraints are listed in Fig.~\ref{fig:all_constraints} ($n_0$ and $E_{SAT}$ are fixed parameters, not constraints) and Table~\ref{tab:constraints}. The availability of so many constraints precludes the dominance of a single experiment or observation or theory and provides data over a wide range of densities. 

The posterior distributions of the EOS calculated from Bayes theorem are shown in Fig.~\ref{fig:nuclear}. 
The 68$\%$ and 95$\%$ CI of the posteriors are represented by the dark green and light blue shaded contours. 
In each case, the posterior uncertainties are much narrower than that of the prior (indicated by the blue dashed curves) and nearly all data fall within the 95$\%$ CI of the posterior regions. The posterior of the density dependence of the symmetry energy shown in Fig. \ref{fig:nuclear}d is similar to that obtained in \cite{decoding2022} but with much narrower uncertainty bands due to the additional astrophysical constraints included in present work. 

In Fig.~\ref{fig:nuclear}f, the purplish blue shaded area shows the EOS for neutron star matter at 90\% CI obtained by employing only astronomical constraints  \cite{legred2021}. 
The uncertainty of our extracted EOS is narrower especially at the low density region where we employ experimental data. At high density the difference in uncertainties may come from the different algorithm used in the analysis. Ref. \cite{legred2021} adopts non-parametric EOSs as priors while the current work uses parametric priors based on an expansion widely used in nuclear physics. The recent constraints \cite{nature2022} obtained by incorporating the $\chi$EFT at low density and Au+Au collective flow data from \cite{russotto2016results, le2016constraining, cozma2018feasibility} are shown as the hatched regions (black dotted lines) corresponding to 68$\%$ (95$\%$) CI. The differences between the current work and Ref. \cite{nature2022} most likely arise from its strict reliance on the $\chi$EFT EOS below 1.5$n_0$ which produces a softer EOS. More comparisons to $\chi$EFT results are discussed below. 

\section{Implications}

\subsection{Composition and the Urca cooling of neutron stars}
Data on the pressure of both symmetric matter and neutron rich matter provides crucial insight into the composition of dense matter. The symmetry energy is very important to understand the composition and dynamics of matter within neutron stars because it contributes to the chemical potentials of the particles that compose the stellar matter. A large symmetry energy may increase the fraction of protons, muons, hyperons or other particles that are present in dense matter. In $\beta$-equilibrium, the proton fraction (protons per baryon) $y_p$ 
increases with increasing symmetry energy. (Details of calculating $y_p$ is given in Methods section).
 
If $y_p$ is large enough, the Urca process of rapid neutrino emission may quickly cool a neutron star \cite{PhysRevLett.66.2701}.  Many isolated neutron stars are observed to cool relatively slowly where the Urca process is likely not operating \cite{Page_2004}. However, some neutron stars, perhaps the more massive ones, are observed to cool quickly consistent with Urca \cite{PhysRevLett.120.182701}. 

\begin{figure}
    \centering
    \includegraphics{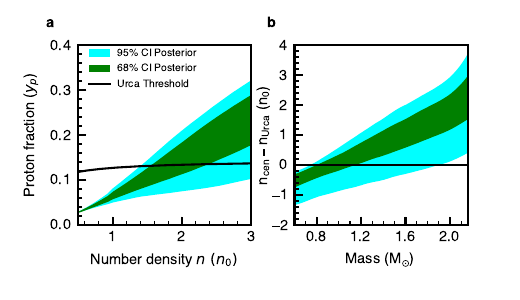}
    \caption{\textbf{Predictions on the proton fraction and the Urca cooling.} The boundaries of the dark green and light blue shaded regions represent 68\% and 95\% CI boundaries of the posterior distributions respectively using an initial sample size of 3 M. (a) The density dependence of the proton fraction. The onset of Urca cooling occurs above the solid black line. (b) The density difference between central density, $n_{cen}$ of a neutron star of mass $M$ and the lowest density, $n_{Ucra}$ where the Urca process is allowed. The y-axis is in unit of saturation density $n_0$.  If this density difference is greater than zero the star may potentially cool quickly.  
\label{fig:URCAden}}   
\end{figure}

In the Urca process a neutron beta-decays followed by electron capture on a proton with the net effect of radiating a $\nu\overline{\nu}$ neutrino pair that cools the star. To conserve both momentum and energy in these weak interactions one needs the Fermi momenta of protons $k_F^p$, neutrons $k_F^n$ and electrons $k_F^e$ to satisfy $k_F^p+k_F^e>k_F^n$. 
At a given density $n$, a symmetry energy above the black solid line in Fig.~\ref{fig:nuclear}e will have a large enough $y_p$ to satisfy this condition so that the Urca process is potentially allowed. Note that the composition, shown in Fig.~\ref{fig:URCAden}a, is a new result made possible by combining symmetric nuclear matter data and symmetry energy data from heavy ion collisions.  

When the central density $n_{cen}$ of a neutron star of mass $M$($M_\odot$) exceeds the Urca threshold density, $n_{Urca}$, the star can potentially cool quickly via the Urca process.
Fig.~\ref{fig:URCAden}b shows the mass dependence of ($n_{cen}-n_{Urca})$, we find that Urca cooling is likely for stars with mass larger than $1.8M_\odot$. Note, neutron star cooling depends on possible super fluid pairing gaps in addition to $y_p$, see for example \cite{10.1111/j.1745-3933.2011.01015.x}. Furthermore, matter in the deep interior of a neutron star may not mainly consist of nucleons. In this simple discussion, we do not include such effects.

\subsection{Benchmarking microscopic theories}

\begin{figure*}
\includegraphics{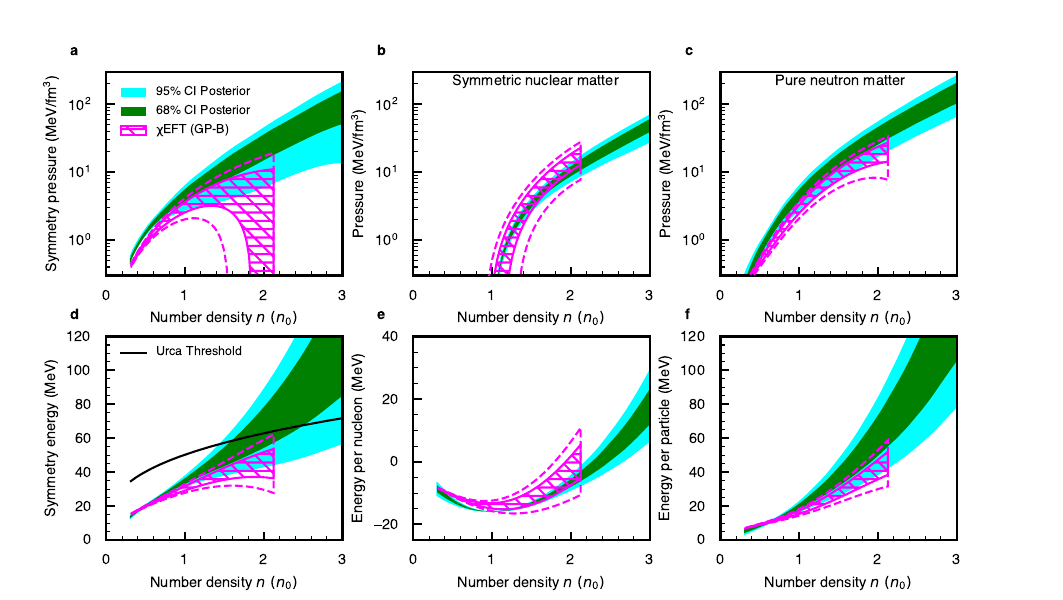}
\vspace*{-8mm}
\caption{\textbf{Energy and pressure of symmetry energy term contribution, symmetric matter and pure neutron matter. } (Top) Left, middle and right plots show the symmetry pressure, pressure of symmetric nuclear matter, pressure of pure neutron matter, respectively, as a function of number density $n$. (Bottom) Same as top panels except the y-axis is energy per nucleon. The solid green (light blue) region is the 68$\%$ (95$\%$) CI of the posteriors using an initial sample size of 3 M. The magenta (hatched) contours represent the calculations from the Chiral Effective Field Theory ($\chi$EFT) \cite{drischler_pnm,drischler_snm,drischler2019}. }
\label{fig:comp_chEFT}
\end{figure*}

The EOS extracted in this work provides the energy per nucleon of symmetric matter, $\epsilon_{SNM}(n)$
and the symmetry energy term, S($n$). The sum of these two terms yields the pure neutron matter EOS $\epsilon_{PNM}(n)$. Unlike previous studies \cite{capano2020,ghosh2022,nature2022}, our methodology does not require input of any theoretical EOS. With recent advances made in the quantification of the uncertainties of the Chiral Effective Field Theory, $\chi$EFT \cite{drischler_pnm}, it has been extended to $2n_0$ and has become a popular choice as the low density EOS to describe neutron star properties \cite{capano2020,ghosh2022,nature2022}. Thus it is interesting to compare our results with the theoretical calculations. In  Fig.~\ref{fig:comp_chEFT},  we show comparisons of the constrained EOS to the Chiral Effective Field Theory ($\chi$EFT) calculations with the 500 MeV momentum cutoff  \cite{drischler2019,drischler_pnm,drischler_snm} for the symmetry energy term of the EOS (left panels), symmetric nuclear matter term (middle panels) and for pure neutron matter (right panels) both in terms of pressure vs. density (upper panels) and energy per nucleon vs. density (lower panels). Note that there are different implementations of $\chi$EFT. The one used here has a truncation that is fully quantified as shown by the solid magenta contours corresponding to 95\% CI and the magenta hatched region corresponding to 68\% CI. 

In general, the extracted EOS is broadly consistent with $\chi$EFT calculations especially at densities below $n_0$. For the symmetric nuclear matter, the agreement is rather good up to 1.5$n_0$ even though it  is slightly  offset from the saturation energy and saturation density as noted in \cite{drischler_snm}. 
 To compare the symmetry term of the EOS, the symmetry energy (pressure) of the $\chi$EFT EOS is obtained by subtracting the energy (pressure) of the symmetric matter from that of the pure neutron matter \cite{drischler2019,drischler_pnm,drischler_snm}. This leads to large uncertainty bands in  Fig.~\ref{fig:comp_chEFT}a and Fig.~\ref{fig:comp_chEFT}d. The EOS derived from $\chi$EFT is in general softer, consistent with smaller predicted NS radii \cite{nature2022}. 
It is interesting to note that due to the softness of the $\chi$EFT, Urca process will be mostly disallowed in NS matter utilizing $\chi$EFT as EOS as shown in  Fig.~\ref{fig:comp_chEFT}d where the black line indicates the Urca threshold.

\section{Summary}
In summary, we perform a Bayesian analyses of 12 nuclear experimental constraints with 3 complementary constraints from astronomical observation to obtain a consistent understanding of the equation of state of neutron rich matter at number densities from about half to three times saturation density.  
Our method eliminates over-reliance on the data of one particular experiment or observation or theory. Within the density range we study, the data do not appear to require the existence of a phase transition inside the core of neutron stars nor do they rule them out. 
We adopt an equation of state constructed from metamodeling in the form of Taylor expansion around saturation density up to fourth order. The extracted values of the expansion coefficients allow construction of the equation of state for symmetric and asymmetric matter including neutron star matter. Even though there are broad agreements between our results and chiral effective theory especially at low density where the theory is more valid, the disagreement increases with density.  The symmetry energy component of the equation of state allows the determination of the thresholds for Urca cooling. With the neutron star equation of state, we obtain a radius of $12.9^{+0.4}_{-0.5}$~km and a deformability of $530^{+115}_{-138}$ for the canonical neutron star with 1.4 solar mass and that heavy mass neutron star cool quickly via Urca process. 
The tension between the posteriors and the data nominally reveals areas where improved measurements are needed. This includes additional measurement of the tidal deformability of the neutron stars and more precise heavy ion data on symmetry energy and pressure above saturation density \cite{sorensen2023dense}.

\section{Methods}

\subsection{Parameterization of the equation of state and the neutron star model}

Neutron star radii and deformabilities combine to provide constraints on matter at densities of $2<n /n_{0}<4$ in the outer core of a cold neutron star ~\cite{abbott2018gw170817,miller2021,riley2021} where we assume the hadronic component of matter is largely nucleons (neutrons and protons). Experimental observables that probe the EOS 
via nucleus-nucleus collisions 
 provide constraints on nucleonic EOS functionals that can be used to describe such densities ~\cite{danielewicz2002determination,le2016constraining,cozma2018feasibility,estee2021probing}. To perform these calculations, we express the dominant hadronic component of the EOS in terms of the energy per nucleon $\epsilon(n,\delta)$ or pressure,
$P=n^{2}\partial{\epsilon}/\partial{n}$, where, $n=n_n+n_p$ is the total density, $n_n$ and $n_p$ are the neutron and proton number densities and $\delta \equiv (n_n-n_p)/n$ is the asymmetry.

To second order in $\delta$, the energy $\epsilon(n,\delta)$ can be written as 
\cite{li2008recent},
\begin{equation} \label{energy}
\epsilon(n,\delta) \approx \epsilon(n,0)+\delta^{2}S(n). 
\end{equation}
\noindent where  $\epsilon(n,0)$, is the energy per nucleon of symmetric matter with equal neutron and proton densities (i.e. $\delta=0$) and  S($n$), is the  
symmetry energy of pure neutron matter where $\delta=1.$ 
We calculate the symmetric matter EOS $(\epsilon(n,0))$ and the density dependence of the symmetry energy (S($n$)) using the Metamodeling ELFc formalism of ref.~\cite{metamodeling2018}.  
In metamodeling, both $\epsilon(n,0)$ and S($n$) contain kinetic energy, (K.E.)  and mean-field potential terms, (U), as well as effective mass ($m^*$) corrections that reflect the non-localities of the nucleonic mean field potentials. These effective mass corrections 
appear in the kinetic energy term where the mass term is replaced by effective masses
\begin{equation}
    \epsilon(n, \delta) = \text{K.E.}(n, \delta, m^*) + U(n, \delta).
\end{equation}

Away from $n=0$ where the kinetic energy has a branch cut singularity, the metamodeling EOS is dominated by the values and derivatives of $\epsilon(n,0)$ and $S(n)$ at saturation density $n_0 \approx \SI{0.16}{nucleons/fm^3}$ which are labeled as follows,

\begin{widetext}
\begin{equation} \label {esat}
E_{SAT} = \epsilon(n_0, 0), K_{SAT}=\frac{1}{2!}\frac{d^2\epsilon(n, 0)}{dx^2}|_{n=n_0}, Q_{SAT}=\frac{1}{3!}\frac{d^3\epsilon(n, 0)}{dx^3}|_{n=n_0}, Z_{SAT}=\frac{1}{4!}\frac{d^4\epsilon(n, 0)}{dx^4}|_{n=n_0},
\end{equation}

\begin{equation} \label {esym}
S_0 = S(n_0), L=\frac{dS(n)}{dx}|_{n=n_0}, K_{sym}=\frac{1}{2!}\frac{d^2S(n)}{dx^2}|_{n=n_0}, Q_{sym}=\frac{1}{3!}\frac{d^3S(n)}{dx^3}|_{n=n_0}, Z_{sym}=\frac{1}{4!}\frac{d^4S(n)}{dx^4}|_{n=n_0},
\end{equation}
\end{widetext}

\noindent where $x = (n-n_0)/(3n_0)$. The exact formulae for these coefficients and how they are combined with additional correction terms to obtain the EOS down to $n=0$ can be found in Ref.~\cite{metamodeling2018}.

By using an EOS of the Meta-modelling form \cite{metamodeling2018}, instead of more customary Skyrme \cite{dutra2012skyrme} or Relativistic Mean Field (RMF) \cite{dutra2014RMF} density functionals, which are commonly used to describe nuclei and nuclear matter, we can reduce the model dependent correlations in asymptotic expansion coefficients listed in Eqs.  (\ref{esat}) and (\ref{esym}) that are characteristic features of the Skyrme and RMF functionals \cite{metamodeling2018}. This allows the  influence of the experimental and astrophysical constraints to be modeled more flexibly and clearly.
Even though the nucleon effective mass $m^*/m_0$ can be an adjustable parameter in the Meta-modeling approach \cite{metamodeling2018}, we assign nominal values of $m^*/m_0= 0.7$ and $m^*_s=m^*_v$ where $m^*_s$ and $m^*_v$ are the isoscalar and isovector effective masses, respectively.

 We adopt  the metamodeling EOS for nuclear matter in the outer core of a neutron star which is composed of neutrons, protons, 
electrons and muons, and is $\beta$-equilibrated at a temperature close to zero MeV. Consistent with our experiment and observation driven approach, we do not introduce a theoretical model for a different form for matter in an inner core of a neutron star, but look to see if strong signatures for its presence can be observed by changes in the EOS at high densities which might emerge in the analyses. We find none. We do ensure when we extend the EOS to higher densities that the speed of sound, $c_s$ (=c 
$\sqrt{\partial P /\partial \epsilon}$) remains less than the speed of light, c. 

To calculate NS radius, we need to extend the EOS from the core to the crust. We adopt a commonly used EOS for the solid outer crust from Ref. \cite{crustalEOS} and extrapolate this through the inner crust region of $0.3n_T <n < n_T$ via spline interpolations. Here, $n_T=-3.75 \times 10^{-4} L+0.0963$ fm$^{-3}$ is the crust to core transition density. In \cite{tsang2020impact} the uncertainty of this crustal EOS description is estimated to be small compared to the present uncertainty of the EOS in the core of the neutron star. Using these EOS expressions, NS properties such as deformability, masses and radii are obtained by solving the Tolman-Oppenheimer-Volkoff (TOV) equations \cite{tolman1939,Oppenheimer1939}. More details of the neutron star model and its uncertainties are provided in \cite{tsang2020impact}.

\subsection{Bayesian Analysis}
\subsubsection{Priors}
Bayesian analysis is used to constrain the relevant Taylor expansion coefficients defined in Eqs.~(\ref{esat}, \ref{esym}). 
Table~\ref{tab:ParRange} 
lists the range of parameters used to generate the priors that will cover most of the experimental and astrophysical constraints listed in 
Table I. 
For comparison, the corresponding posterior values are also listed in Table~\ref{tab:ParRange}.
Here, we give the rationale in choosing the ranges of the parameters listed in Table~\ref{tab:ParRange} using experimental information as 
guidance when possible. The general principle is to allow each parameter with a very wide range to ensure that the priors encompass as much of the experimental 
constraints listed in Table I as possible.

\begin{table*}
\caption{\textbf{Priors and posterior values for various EOS parameters and predictions.} (Top) Ranges of Taylor expansion coefficients used to generate priors and posteriors. (Bottom) Predictions for Taylor expansion parameters at $n=\SI{0.1}{fm^{-3}}$. The parameters of the first group, $K_\text{SAT}$, $Q_\text{SAT}$, $S_\text{0}$, 
$L$, $K_\text{sym}$, $Q_\text{sym}$, $Z_\text{sym}$, and $P_\text{SNM}(4n_0)$ are varied uniformly within the ranges listed under priors. The symmetric matter constraint at supra-saturation density dictates that $Z_\text{SAT}$ values are strongly correlated with $Q_\text{SAT}$. 
} \vspace{0.2in}
\label{tab:ParRange}
\begin{tabular*}{\linewidth}{@{\extracolsep{\fill}}lcc}
\hline
\hline
      Parameters &  Priors & Posteriors \\
      & &  (Median $\pm$ 68\% CI)\\
\hline
$K_\text{SAT}$ (MeV)  & [0, 648]  & $221^{+23}_{-27}$ \\
$Q_\text{SAT}$ (MeV) & [-1100, 2100] & $-640^{+273}_{-249}$ \\
$S_\text{0}$  (MeV) &   [24.7, 40.3]     & $34.9^{+1.7}_{-2.0}$ \\
$L$  (MeV) &       [-11.4, 149.4]          & $83.6^{+19.2}_{-16.8}$ \\
$K_\text{sym}$  (MeV) &   [-328.5, 237.9]   & $-6^{+102}_{-102}$ \\
$Q_\text{sym}$ (MeV) & [-489, 1223] & $692^{+377}_{-548}$ \\
$Z_\text{sym}$ (MeV) & [-10110, 2130] & $-149^{+1585}_{-1982}$ \\
$P_\text{SNM}(4n_0)$ (MeV/fm$^{3}$) & [0, 300] & $213^{+60}_{-66}$ \\
\hline
\hline
      Predictions & &  Posteriors \\
   &  &  (Median $\pm$ 68\% CI) \\
\hline
$R(1.4M_\odot)$ (km)  & & $12.9^{+0.4}_{-0.5}$ \\
$\Lambda(1.4M_\odot)$ & & $530^{+115}_{-138}$ \\
$S_\text{01}$  (MeV)  & & $24.0^{+0.5}_{-0.5}$ \\
$L_\text{01}$  (MeV)  & & $57^{+4}_{-6}$ \\
$K_\text{01}$  (MeV)  & & $-35^{+42}_{-42}$ \\
$Q_\text{01}$  (MeV)  & &  $200^{+115}_{-148}$\\
$Z_\text{01}$  (MeV)  & &  $-528^{+335}_{-335}$\\

\hline
\end{tabular*}
\end{table*}

 The ranges of the symmetry energy parameters of Eq.~\eqref{esym} are guided by the comprehensive study in Ref. \cite{decoding2022}. We do not have a corresponding comprehensive analysis for the symmetric matter EOS parameters of Eq.~\eqref{esat}. Together with the well-known values of $n_0$, $E_{SAT}$ and $K_{SAT}$, the existence of two consistent constraints at $2n_0$ on the symmetric matter EOS provides some guidance on the EOS up to $2n_0$. 

Analyses of recent experimental results at high collision energy and therefore high density from the Hades collaboration ~\cite{adamczewski2023proton} and from the Beam Energy Scan Collaboration \cite{star2021,star2022} have revealed disagreements between the new data and data from E895 Collaboration \cite{E895_1999,E895_2000}. The latter was used to derive the HIC(DLL) contours. The new data, has stimulated a re-evaluation of the EOS for symmetric matter \cite{Oliinychenko:2022uvy} at high density, which may lead to a future reevaluation of the pressure constraints currently provided by HIC(DLL) at densities greater than $3n_0$ . For the prior distributions of the symmetric matter EoS in this work, we have assessed both the existing HIC(DLL)  and new but still preliminary analysis \cite{Oliinychenko:2022uvy} and concluded that the symmetric matter pressure at $4n_0$ does not exceed $\SI{300}{MeV/fm^{3}}$. We include this knowledge in the prior distribution by selecting values for $Z_{SAT}$ so that the symmetry pressure $P_\text{SNM}(4n_0)$ is uniformly distributed from $0 - \SI{300}{MeV/fm^{3}}$, nearly twice the published HIC(DLL) standard deviation upper limit of $\SI{156}{MeV/fm^{3}}$ \cite{danielewicz2002determination}.  Effectively, this means that $Z_{SAT}$ prior is not independent but is correlated with the priors for other Taylor coefficients, especially $Q_{SAT}$, in the expression for the symmetric matter EOS. Thus, $P_\text{SNM}(4n_0)$ appears in Table~\ref{tab:ParRange} instead of $Z_{SAT}$. The current prior distributions are wide enough to encompass possible higher pressures from the preliminary analysis shown in \cite{sorensen2023dense, Oliinychenko:2022uvy}   

 \begin{figure}
    \begin{center}
        \includegraphics[width=\linewidth]{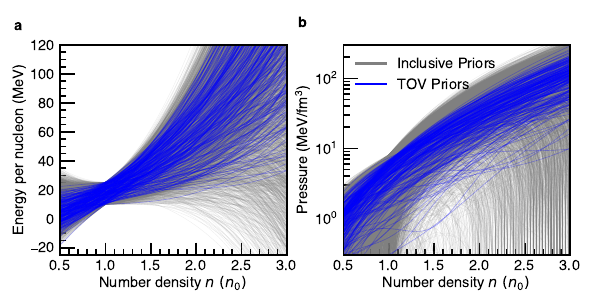}
        \end{center}
        \vspace*{-5mm}
\caption{\textbf{Prior EOS for pure neutron matter and priors that form stable neutron stars with maximum neutron star mass greater than $2.17M_{\odot}$}. The former is shown as the grey curves while the latter is shown as the blue curves. The y-axis is energy per nucleon in the left plot and pressure in the right plot.
\label{fig:priors}}
\end{figure}

To illustrate the influence of neutron star on the priors, 5000 randomly chosen priors based on the parameter ranges listed in Table~\ref{tab:ParRange} are shown as grey curves in Fig.~\ref{fig:priors} 1 where the left panel shows the density dependence of energy per particle of the pure neutron matter (commonly used in nuclear physics community) while the right panels shows the corresponding pressure as a function of number density (more commonly used in astrophysics community). The TOV equations are solved and any EOS which do not lead to a maximum neutron star mass of at least $2.17M_{\odot}$  are rejected before applying the Bayesian analysis. (Lowering the least maximum neutron star mass to $1.8M_{\odot}$ does not change the results/conclusions significantly.) The neutron star requirement removes very stiff and very soft EOS especially those with very small or negative pressure at high density. This greatly reduces the number of priors to about 5\% of the original numbers. They are shown as blue curves in Fig.~\ref{fig:priors}. Since these are the priors used in the Bayesian analysis, they will be referred to simply as ``priors" from here on. 
\begin{figure*}
    \centering
    \vspace*{-3mm}
    \includegraphics{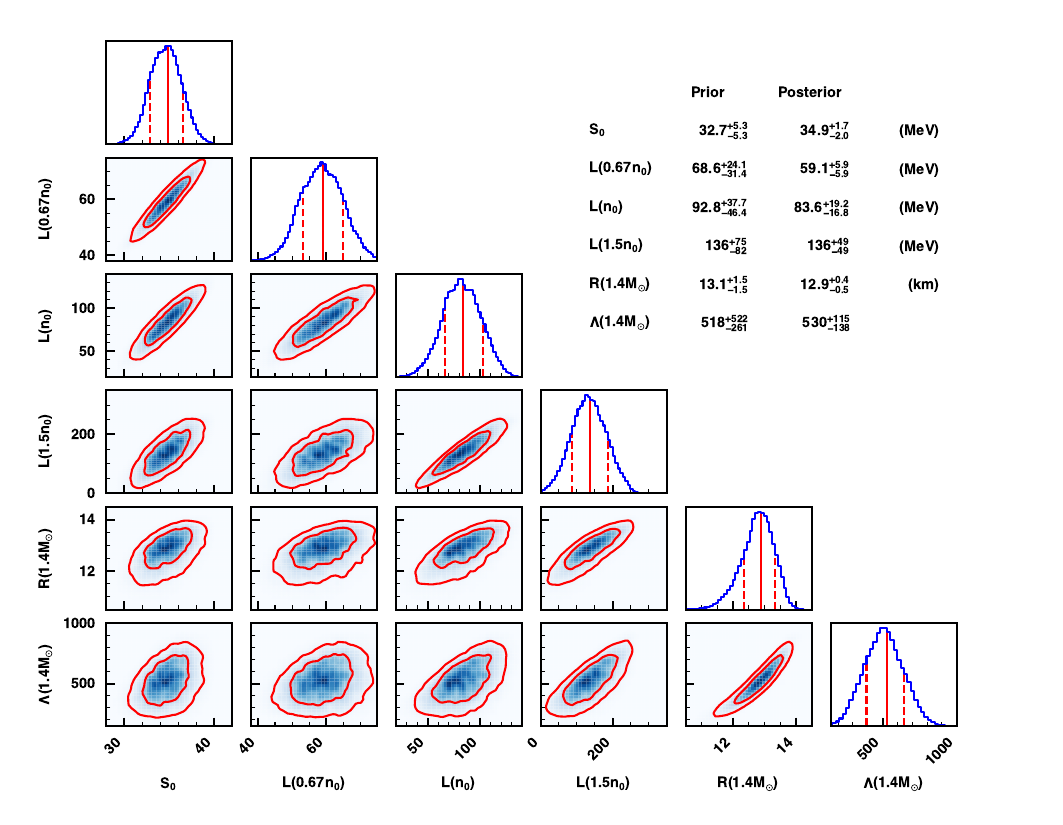}
    \vspace*{-10mm}
    \caption{\textbf{Marginalized probability distributions for EOS parameters.} The off-diagonal plots show the pair-wise probability distributions for $S_0$, $L(0.67n_0)$, $L(n_0)$, $L(1.5n_0)$, $R(1.4M_\odot)$ and $\Lambda(1.4M_\odot)$. The two red contours, from inside to outside, correspond to 68\% and 95\% confidence intervals, respectively. The diagonal plots show the marginalized probability distribution of each parameter correspond to posterior after all experimental constraints are applied. The two outer red vertical dashed line shows the 68\% confidence interval and the single solid red vertical line shows the median (50$^\text{th}$ percentile). The table on the upper right corner of the figure shows the median $\pm$ 68\% confidence interval values.}
    \label{fig:cornerplot}
\end{figure*}

\subsubsection{Bayes theorem}

Bayesian inference is performed to generate the posterior distribution, which is governed by various expansion coefficients defined in Eqs.~(\ref{esat}, \ref{esym}) and the experimental and astronomical measurements. The posterior distribution is given by Bayes theorem \cite{bayes_theorem}, which can be written as,

\begin{equation}
    P(\text{EOS}|\text{Constraints}) \propto P(\text{EOS})\prod_i \mathcal{L}(i^\text{th}\text{ Constraint}).
\label{eq:bay}
\end{equation}

In this equation, $P(\text{EOS})$ is the prior and $\mathcal{L}(i^\text{th}\text{ Constraint})$ is the likelihood, defined as the probability of observing the experimental and astronomical results assuming that a given set of EOS expansion coefficients is the perfect description of nuclear matter. In this analysis, for an observable $O$, which the measurements/observations constrain to have a mean of $x_0$ and a standard deviation $\sigma$ equal to the error, the likelihood of it having a true value of $x$ is given by a Gaussian $\exp[- (x -x_0)^2/(2 \sigma^2) ]$. 

For gravitational wave constraints with asymmetric uncertainty, an asymmetric Gaussian is used. The latter is constructed from two Gaussian functions, having the same mean value but different widths and scaled to match the height of each other at the mean. These are used to describe the two halves of the distribution that lie on opposite sides of the mean value. 

\subsubsection{Posterior distributions}

The Bayesian analysis is performed with parameters allowed to vary within ranges listed in Table~\ref{tab:ParRange}. The posterior of EOS parameters and predictions for the radius and deformability of the 1.4$M_\odot$ neutron star are also listed. In the same table, we listed the predictions for the Taylor expansion coefficients for the symmetry energy at the density of $\SI{0.1}{fm^{-3}}$ where the experimental constraints are most robust. Parameter is defined as variables that EOS takes as input and prediction is defined as everything else that requires calculation to be performed to get the values. We made this distinction for the tables for clarity, but they will all be called parameters in other sections for brevity. 

The posterior distribution, which describes the probability of the $i^{th}$ parameter having value $m_i$ given the constraints listed in Table I, is calculated from Bayes Theorem.
The marginalized posterior distributions are calculated by distributing each parameter on a histogram, weighted by a factor from Eq.~\eqref{eq:bay}. A total of 
4,500,000 EOSs were sampled in two stages. In the first stage, 1,500,000 EOS were  sampled uniformly within ranges in Table~\ref{tab:ParRange}. This small set of 
expansion coefficients informed us of the range of preferred values. The remaining 3,000,000 expansion coefficient sets were sampled uniformly within 99\% confidence intervals (CIs)
from the marginalized distributions from the initial batch of EOS . This two-step process allows us to more efficiently 
sample the parameter space that is relevant. The CIs for the energy and pressure values for the EOS, as well as the values of various observables 
directly related to the nuclear EOS, were calculated by weighing each prediction with the corresponding factor. 

Figure~\ref{fig:cornerplot} shows
the corner correlation plots for six parameters: symmetry energy at saturation density, $S_0$, slopes of the symmetry energy at 0.67$n_0$, $n_0$ and 1.5$n_0$, and the radii and deformability of a
neutron star with the nominal mass of 1.4 $M_\odot$. In red, we show the $1\sigma$ and $2\sigma$ contours of these correlations. Because we construct the
prior EOS using meta-modeling, we avoid any a priori correlations between the slope of the symmetry energy at different densities and other
 neutron star properties. The correlations in these plots reflect strong connections between the EOS at specific densities and neutron star radii and deformabilities. Nonetheless, both the radii and deformability of the neutron star are correlated with the slope of the symmetry energy at
0.67$n_0$, $n_0$ and 1.5$n_0$ as well as with each other, consistent with 
previous studies. Along the diagonal are the posterior 1D plots for the parameters with the solid vertical lines corresponding to the median (50$^\text{th}$ percentile) and the dashed vertical lines corresponding to the 68\% ($1\sigma$) CI. 
The posterior values with $1\sigma$ uncertainties are also listed in Table~\ref{tab:ParRange}. In each case, the posterior uncertainties are much narrower than that of the prior. 

\subsection{Direct Urca cooling}

The symmetry energy is very important to understand the composition and dynamics of matter within neutron stars because it contributes to the chemical potentials of the particles that compose the stellar matter. In $\beta$-equilibrium the chemical potentials of neutron $\mu_n$, proton $\mu_p$ and electron  $\mu_e$ satisfy $\mu_e=\mu_n-\mu_p$ and the proton fraction $y_p$ satisfies,
\begin{equation}
   [4S(n)(1-2y_p)]^3+\bigl\{[4S(n)(1-2y_p)]^2-m_\mu^2\bigr\}^{3/2}=3\pi^2ny_p, 
\end{equation}
at baryon density $n$ with $m_\mu$ being the muon mass.

\section*{Data availability}
All data points used in this work are listed in Table 1 together with the references. They are also plotted in Fig. 2. 

\section*{Code availability}
Analysis codes specially written for this work are available online at https://github.com/nscl-hira/TidalPolarizabilityPublic. The TOV solver used in this work is available upon request to the corresponding author.

\section*{Acknowledgments}
The authors would like to acknowledge many stimulating discussions with the participants at the workshops sponsored by the Institute of Nuclear Theory in 2021 and 2022 and with members of the Transport Model Evaluation Project. The authors are grateful to Prof. Christian Drischler for providing the $\chi$EFT calculations shown in our figures. This work is supported in part by the National Science Foundation under Grant No. PHY-2209145 (RK, WJG, CYT, MBT) and the US Department of Energy Office of Science Office of Nuclear Physics grant no. DE-FG02-87ER40365 (CJH). 

The Facility for Radioactive Ion Beams funded by the US Department of Energy, is committed to fostering a safe, diverse, equitable, and inclusive work and research environment which values respect and personal integrity. The authors adhere to the FRIB research code of conduct in accordance with the highest scientific, professional, and ethical standards as detailed in https://frib.msu.edu/users/pac/conduct.html. 

In an ideal world, it should not be necessary to identify the authors by gender or from under-represented group. Until the ideal world is reached, the authors acknowledge that our references and citations most likely under-represent contributions from women and minorities. We futher acknowledge that Michigan State University occupies the ancestral, traditional and contemporary Lands of the Anishinaabeg – Three Fires Confederacy of Ojibwe, Odawa and Potawatomi peoples. In particular, the university resides on Land ceded in the 1819 Treaty of Saginaw.

\section*{Contributions}
CYT and MBT first conceived the project. CYT wrote the software to do the Bayesian Analysis. CYT, MBT and WGL collected and evaluated all the constraints used in the analysis. RK researched, reviewed and validated the final set of constraints adopted in Table 1. RK also wrote some of the software codes to analyze the Bayesian results and generated all the figures with data.  CJH contributed to the impact of Urca cooling section. All authors contributed in writing, editing and revising the manuscript. The ordering of the authors reflects the length of time the authors have joined the project. 

\section*{Competing Interests}
The authors declare no competing financial or non-financial interests.

\end{document}